\documentclass{emulateapj}
\usepackage{natbib}
\usepackage{epsfig}
\usepackage{graphicx}
\usepackage{subfigure}
\usepackage{float}
\usepackage{amsmath}
\usepackage{color}
\usepackage{amssymb}
\usepackage{amsfonts}
\usepackage[colorlinks,linkcolor=blue,anchorcolor=green,citecolor=blue]{hyperref}
\usepackage{amssymb,amsmath}
\usepackage{upgreek}
\usepackage{color}
\usepackage[normalem]{ulem}
\usepackage{verbatim}
\newcommand{\degree}{\ensuremath{^\circ}}
\newcommand{\hs}{\ensuremath{^{\mathrm{s}}}}
\newcommand{\hm}{\ensuremath{^{\mathrm{m}}}}
\newcommand{\hh}{\ensuremath{^{\mathrm{h}}}}
\def\be{\begin{equation}}
\def\ee{\end{equation}}
\shorttitle{FRB 121102 : Constraints from the 8-year Fermi-LAT Data}
\shortauthors{Zhang \& Zhang}
\begin{document}

\title{Repeating FRB 121102 : Eight-year Fermi-LAT Upper Limits and Implications}

\author{Bin-Bin Zhang\altaffilmark{1,2}, Bing Zhang\altaffilmark{3}}

\affil{$^1$Instituto de Astrof\'isica de Andaluc\'a (IAA-CSIC), P.O. Box 03004, E-18080 Granada, Spain; zhang.grb@gmail.com \\
$^2$Scientist Support LLC, Madsion, AL 35758, USA \\
$^3$Department of Physics and Astronomy, University of
Nevada, Las Vegas, NV 89154, USA; zhang@physics.unlv.edu }

\begin{abstract}
The repeating fast radio burst (FRB) source that produced FRB 121102 was recently localized in a star forming galaxy at $z=0.193$, which is associated with an extended radio source at the burst location. One possibility is that the repeating FRBs are produced by a new-born magnetar, which also powers the radio nebula. If so, the magnetar may produce $\gamma$-ray emission due to magnetic dipolar spin-down. The luminosity depends on the magnetar spin parameters and age. We process the eight-year Fermi LAT data at the position of FRB 121102 and place an energy flux upper limit of $\sim 10^{-11} \ {\rm erg \ cm^{-2} \ s^{-1}}$ in time bins with six-month intervals, and an accumulated energy flux upper limit of $\sim 4\times 10^{-12} \ {\rm erg \ cm^{-2} \ s^{-1}}$ over the eight year span. The corresponding $\gamma$-ray luminosity upper limits are $\sim 10^{45}  \  {\rm erg \ s^{-1}}$ and $\sim 4\times 10^{44} \ {\rm erg \ s^{-1}}$ for the time-resolved and time-integrated analyses, respectively. We discuss the implications of these limits on the young magnetar model.

\end{abstract}


\section{Introduction}
\label{sec:intro}

The breakthrough discovery of the host galaxy of the repeating fast radio burst (FRB) source that produced FRB 121102 \citep{chatterjee17,marcote17,tendulkar17} confirmed the {cosmological origin of at least one of the FRBs} \citep{lorimer07,thornton13,petroff15} and opened the possibility of seriously considering the progenitor models of these mysterious sources. The star forming host galaxy of FRB 121102 \citep{tendulkar17} at  $z=0.193$ suggests a possible connection between FRBs and deaths of massive stars. One possibility is that the source is a new-born millisecond magnetar which powers the FRBs \citep{murase16,metzger17} and also the radio nebula \citep{yang16,dai17}, with the birth of the pulsar probably associated with a GRB or an ultra-luminous supernova \citep{metzger17}. Within this picture, the new-born magnetar needs to be spinning rapidly to have a large enough spin-down luminosity in order to power the repeating FRBs. Several estimates of such requirements have been proposed recently \citep[e.g.][]{metzger17,kashiyama17,cao17}. 

The large spin-down power of the new-born magnetar would produce $\gamma$-ray emission due to magnetospheric activities and interactions between the pulsar wind and the surrounding medium (likely the nebula itself). It would be then interesting to search for possible $\gamma$-ray signals from the source using the archival Fermi Large Area Telescope (LAT) data. This paper reports the results from such a search.

\section{The Fermi/LAT Observation of FRB 121102}
\label{sec:obser}

\begin{figure}

\begin{tabular}{c}
\includegraphics[keepaspectratio, clip, width=0.45\textwidth]{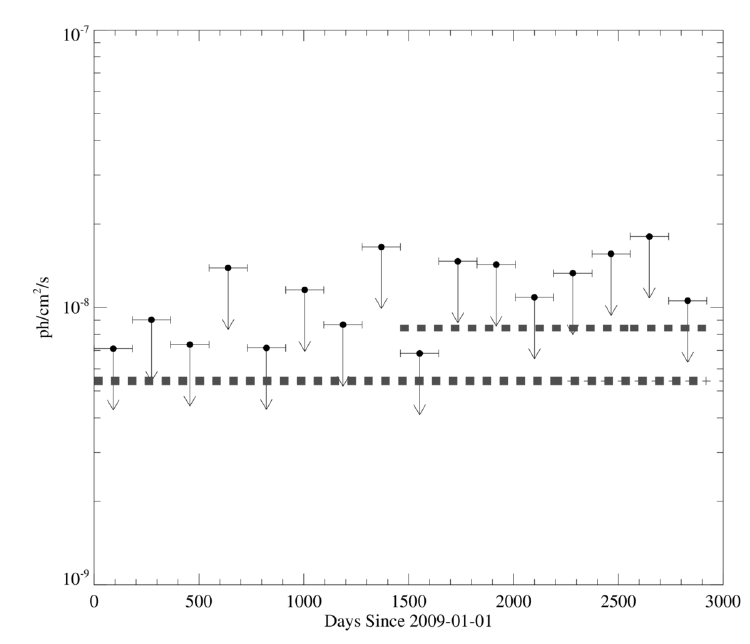} \\
\end{tabular}

\caption{The photon flux upper limits of 100 MeV-10 GeV LAT band emission of FRB 121102 in the 8-year time span. Each down arrow denotes flux limit for each 6-month time bin; the short and long dashed lines represent the 4-year and 8-year up-limits, respectively. }
\label{fig:fig1}
\end{figure}

We extract the photon files and the spacecraft history files from LAT data server\footnote{\url{https://fermi.gsfc.nasa.gov/cgi-bin/ssc/LAT/LATDataQuery.cgi}} around the average J2000 position of FRB 121102 - RA $ = 05\hh31\hm58.70\hs$,  Dec$ = +33\degree08\arcmin52.5\arcsec$ \citep{chatterjee17}, with a time span between 2009-01-01 and 2016-12-31, an energy range of 100 MeV - 10 GeV, and a spatial radius of 60 degree. We then process the LAT data using the standard Fermi Science Tools (v10r0p5). There was no $\gamma$-ray transient found in the Fermi/LAT data in a previous research \citep[e.g.,][]{2016ApJ...833..177S}, and no known source is available in the Fermi/LAT 4-year Point Source Catalog \citep[3FGL;][]{2015ApJS..218...23A} at the position. We therefore only focus on the flux upper limits in this work. We divide the time range into 16 bins, each containing six-month data. We analyze the 100 MeV - 10 GeV photons within 10 degrees around the FRB 121102 location for each time bin, for a four-year time span from 2012-12-31 to 2016-12-31, and for the entire eight-year span. {We selected the ``Pass 8 Source" class (evclass = 128, evtype = 3)  LAT photons with the contraint that the region of interest (ROI) does not go below the gamma-ray-bright Earth limb (defined at 105 degree from the Zenith angle).  }
Diffuse gamma-ray emission from the Milky Way is estimated using the official Galactic interstellar emission model {\tt gll\_iem\_06.fit}. The isotropic diffuse component is treated with the Fermi official isotropic spectral template {\tt iso\_P8R2\_SOURCE\_V6\_v06.txt}. All individual objects appearing in the Fermi 3FGL catalog \citep{2015ApJS..218...23A} within a $10^{\circ}$ radius of the FRB position are
included in the sky model as separately-fit point sources with fixed positions. No significant (TS $<$ 25) source is detected in any of the time bins. We thus calculate the photon flux upper limit at 95\% confidence level using the ``integral" method included in the Fermi Science Tools. This method calculates the upper limit by integrating the likelihood function to the given 95\% level \citep{Roe-Woodroofe,Feldman-Cousins}. A power-law spectrum model with photon index =$-2$  is assumed. The results of those upper limits, together with the calculated upper limits of the energy flux and {isotropic} luminosity, are listed in Table 1. The photon flux and luminosity upper limits are also plotted in Figure \ref{fig:fig1}.

\begin{table}
\caption{Upper limits of LAT observations of FRB 121102}

\begin{tabular}{cclll}
\hline
t1 & t2& Photon Flux  & Energy Flux  & Luminosity \\ 
 &  & (ph/cm$^2$/s)  &(erg/cm$^2$/s) &(erg/s) \\

\hline

2009-01-01&2009-07-02&                               7.10$\times 10^{-9}$&                               5.30$\times 10^{-12}$&                               5.60$\times 10^{44}$\\
2009-07-02&2010-01-01&                               9.02$\times 10^{-9}$&                               6.72$\times 10^{-12}$&                               7.11$\times 10^{44}$\\
2010-01-01&2010-07-02&                               7.34$\times 10^{-9}$&                               5.47$\times 10^{-12}$&                               5.79$\times 10^{44}$\\
2010-07-02&2011-01-01&                               1.39$\times 10^{-8}$&                               1.03$\times 10^{-11}$&                               1.09$\times 10^{45}$\\
2011-01-01&2011-07-03&                               7.15$\times 10^{-9}$&                               5.33$\times 10^{-12}$&                               5.63$\times 10^{44}$\\
2011-07-03&2012-01-01&                               1.16$\times 10^{-8}$&                               8.62$\times 10^{-12}$&                               9.12$\times 10^{44}$\\
2012-01-01&2012-07-02&                               8.66$\times 10^{-9}$&                               6.46$\times 10^{-12}$&                               6.83$\times 10^{44}$\\
2012-07-02&2012-12-31&                               1.65$\times 10^{-8}$&                               1.23$\times 10^{-11}$&                               1.30$\times 10^{45}$\\
2012-12-31&2013-07-02&                               6.83$\times 10^{-9}$&                               5.09$\times 10^{-12}$&                               5.38$\times 10^{44}$\\
2013-07-02&2014-01-01&                               1.47$\times 10^{-8}$&                               1.09$\times 10^{-11}$&                               1.16$\times 10^{45}$\\
2014-01-01&2014-07-02&                               1.43$\times 10^{-8}$&                               1.06$\times 10^{-11}$&                               1.12$\times 10^{45}$\\
2014-07-02&2015-01-01&                               1.09$\times 10^{-8}$&                               8.11$\times 10^{-12}$&                               8.58$\times 10^{44}$\\
2015-01-01&2015-07-03&                               1.33$\times 10^{-8}$&                               9.90$\times 10^{-12}$&                               1.05$\times 10^{45}$\\
2015-07-03&2016-01-01&                               1.56$\times 10^{-8}$&                               1.16$\times 10^{-11}$&                               1.23$\times 10^{45}$\\
2016-01-01&2016-07-02&                               1.80$\times 10^{-8}$&                               1.34$\times 10^{-11}$&                               1.42$\times 10^{45}$\\
2016-07-02&2016-12-31&                               1.06$\times 10^{-8}$&                               7.88$\times 10^{-12}$&                               8.34$\times 10^{44}$\\
\hline
2012-12-31&2016-12-31&                               7.66$\times 10^{-9}$&                               5.71$\times 10^{-12}$&                               6.04$\times 10^{44}$\\
2009-01-01&2016-12-31&                               8.43$\times 10^{-9}$&                               4.05$\times 10^{-12}$&                               4.28$\times 10^{44}$\\

\hline

\end{tabular}
\end{table}

\section{Implications of the upper limits}

In order to connect the luminosity upper limits with the magnetar parameters, we adopt the standard dipole spin-down formulae for pulsars
\citep[e.g.][]{shapiro83,zhangmeszaros01} 
\be
L_{\rm max}=\frac{L_{0}}{(1+T/{T}_{0})^2},
\label{eq:l}
\ee
where $L_{0}$ and ${T}_{0}$ are the characteristic luminosity and the timescale, i.e.
\be
L_{0}=\frac{I\Omega_0^2}{2{T}_{0}}\simeq 1.0\times10^{45}
{\rm erg~s^{-1}}B_{p,13}^2P_{0,-3}^{-4}R_6^6,
\label{Leem0}
\ee
\be
{T}_{0}=\frac{3c^3I}{B_p^2R^6\Omega_0^2} \simeq 0.65~{\rm
yr}~ I_{45}B_{p,13}^{-2}P_{0,-3}^2R_6^{-6}.
\label{Teem}
\ee
Here $\Omega_0$ is the initial angular velocity when the pulsar was born, $B_{p}$
is the surface dipolar magnetic field strength at the pole, $P_0 = 2\pi / \Omega_0$ 
is the initial rotation period, $I$ is the moment of inertia, $R$ is the neutron star radius, and the convention $Q = 10^n Q_n$ has been adopted in cgs units.

The LAT-band $\gamma$-ray luminosity may be related to the spin-down luminosity through a parameter, i.e.
\be
L_\gamma = \eta L_{\rm sd},
\ee
where $\eta = \eta_\gamma f_b^{-1}$, $\eta_\gamma < 1$ is the efficiency parameter (the fraction of spin-down energy that goes to $\gamma$-rays), and $f_b = \Delta\Omega/4\pi $ is the beaming factor. The $\eta$ parameter can be either greater or less than unity depends on the values of $\eta_\gamma$ and $f_b$. {The LAT-band emission may be either produced directly from the pulsar magnetosphere, or through dissipation of the pulsar wind outside the light cylinder. For the first possibility, there are arguments that near-surface GeV $\gamma$-rays are attenuated via pair production or photon splitting \citep[e.g.][]{zhang01,beloborodov13}, so that magnetars may not be bright LAT-band emitters. LAT band emission may come from the outer gaps \citep[e.g.][]{2001ApJ...562..918C,2002ApJ...579..716Z}. However, the efficiency of converting the spindown luminosity to $\gamma$-ray luminosity may not be high, as suggested by the Fermi-LAT upper limits on emission from known magnetars \citep[e.g.,][]{2010ApJ...725L..73A,2017ApJ...835...30L}. As a result, we envisage that bright $\gamma$-ray emission may be generated outside the light cylinder through direct dissipation via shocks or reconnection.}

The age $T$ of the magnetar is not known. The magnetar may be born before Fermi was launched in 2008. In that case, the age is longer than 8 years. The magnetar may be also born after Fermi observations started but before the first FRB burst was detected. In this case, one can in principle observe the magnetar from $T \sim 0$. For the scenario of producing a young magnetar associated with the death of a massive star, the magnetar may be initially surrounded by a heavy supernova ejecta so that $\gamma$-rays are trapped and cannot be directly observed. In this case, $T$ should correspond to the age when $\gamma$-rays become transparent.

There are cases when the $\gamma$-ray emission of the new born millisecond magnetar would be directly detected shortly after birth. For the magnetar birth from a core collapse supernova, a relativistic jet may be launched which would be manifested as a gamma-ray burst (GRB). If the jet beams towards Earth, there would be a GRB which is associated with a bright afterglow. Our tight upper limits essentially rule out such a possibility. A second possibility is that a massive millisecond magnetar may be formed due to NS-NS mergers \citep{dai06,fan06,metzger08,zhang13}, which would not be buried by a heavy envelope. The remnant may become transparent within $T_{\rm trans} \sim 10^3$ s \citep[e.g.][]{sun17}, so that $\gamma$-ray emission from the magnetar may be directly detected. In this case, $T$ may be adopted as a small value $T_{\rm trans}$.

Our observations require that the maximum $\gamma$-ray luminosity at the relevant age $T$ is smaller than the upper limits, i.e.
\be
L_{\rm \gamma}(T) \le L_{\rm lim}(T),
\ee 
{where $L_{\rm lim}(T) = L_{\rm lim}  (8 {\rm yr}/T)^{1/2}$, and $L_{\rm lim} = 4.28\times 10^{44} ~{\rm erg~s^{-1}}$ is the eight-year limit presented in the last row last column in Table 1.}

Given a set of magnetar parameters $P_0$ and $B_p$, the spin-down luminosity depends on the age of the magnetar, $T$. Since the earliest burst, FRB 121102, was detected in late 2012, the magnetar needs to be at least 4 years old until now. We consider two possibilities. 

\begin{figure}

\begin{tabular}{c}
\includegraphics[keepaspectratio, clip, width=0.45\textwidth]{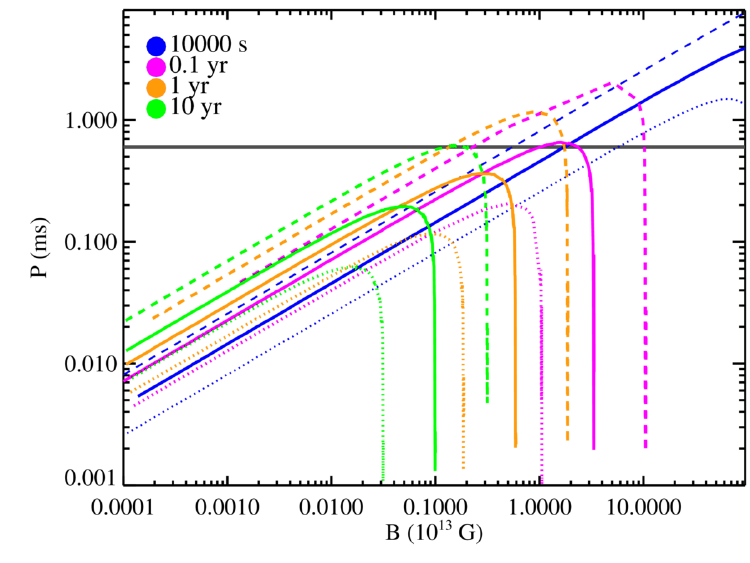} \\
\end{tabular}
\caption{Constraints of the pulsar parameter by the LAT upper limit.  {\it Solid lines}: cases of $\eta=1$; {\it Dotted lines}: cases of $\eta=0.1$; {\it Dashed lines}: cases of $\eta=10$. The horizontal line denotes the break-up limit $P_0=0.6$ms. The  parameters are constrained to the regions which are above the curves/lines.  }
\label{fig:fig2}
\end{figure}

First, if the pulsar was born between 2009-01-01 and 2012-11-02, one should use the data during that period of time to constrain magnetar parameters assuming different transparent times $T_{\rm trans}$ by setting $T = T_{\rm trans}$ in Eq.(\ref{eq:l}). The transparent time depends on the amount of ejecta and opacity, which may range from minutes (for NS-NS mergers) to $\sim$ yr (for supernovae). In our calculation, we use $10^2$ s, $10^4$ s, 0.1 yr, 0.5 yr,
and 1 yr, respectively. Since the upper limits are slightly different in different time bins, we use the average upper limits to derive the constraints as presented in Fig. \ref{fig:fig2}.

The second possibility is that the magnetar was born before 2009-01-01. For these cases, the time $T$ should be adopted as the age of the magnetar when the first observation started (in 2009-01-01). When the birth time is much earlier than 2009-01-01, one can use the tighter upper limit over a much longer time (e.g. 4-year or 8-year) limit to perform the constraint. For this case, we consider $T=10$ yr as an example.

The results for $T= 0, 10^2$ s, $10^4$ s, $0.1$ yr, 0.5 yr, $1.0$ yr, and 10 yr are plotted in the
$(B_p, P_0)$ plane ($I_{45}=1$ and $R_6=1$ assumed) as shown in Figure \ref{fig:fig2}. 
For each $T$, we consider $\eta = 0.1$, $\eta=1$ and  $\eta=10$. The region above each curve is the parameter space allowed for the given $T$ and $\eta$ value. Since the minimum spin period of a millisecond pulsar would be the break-up limit $P_{\rm 0,min} \sim 0.6$ ms \citep[e.g.,][]{2006MNRAS.370L..14V}, our results suggest that there is essentially no constraint from the upper limits if the pulsar age $T > 0.1$ yr in the case of $\eta =1$. This means that the current limit cannot constrain the magnetar parameter if it was born before Fermi observations started. On the other hand, for $\eta=1$ and if the magnetar age is 4-8 yr and the transparent time shorter than 0.1 yr, the upper limits in the first four years could place some constraints on $B_p$ of the pulsar. In particular, if the transparent time is short enough, the non-detections exclude high magnetic field strengths. As a result, if FRB 121102 progenitor was indeed born after 2009-01-01 and has a magnetar origin, it must have been born in a dirty environment with a delayed transparent time (say $T_{\rm trans} > 0.1$ yr) in order to evade detections. We note that above constraints can be tighten if $\eta >1$ (e.g, $\eta =10$, dashed line in Figure 2).

\section{Summary and Discussions}
\label{sec:summary}

We have placed some 100 MeV - 10 GeV luminosity upper limits on the FRB 121102 source, which are about $10^{45} \ {\rm erg \ s^{-1}}$ for six-month time bins and $\sim 4 \times 10^{44} \ {\rm erg \ s^{-1}}$ for the entire eight-year span. The upper limits do not pose significant constraints on the young magnetar parameters. In particular, there is no constraint if the young pulsar/magnetar was born before 2009-01-01, or if it was born between 2009-01-01 and 2012-11-02 but the magnetar transparent time is longer than 0.1 yr. However, if the transparent time is short, the non-detections suggest that the dipole field strength cannot be too high (for $\eta=1$). {The limits also have no constraints on other models that do not invoke a magnetar as the source of the repeating FRBs \citep[e.g.][]{cordes16,connor16,2016ApJ...829...27D,zhang17}.}

Since the photon flux upper limits of multi-year Fermi LAT observations for other sky regions are of this order, the luminosity upper limits for other FRBs would be of the same order unless the source is much closer. As a result, more significant constraints on the FRB source using Fermi/LAT $\gamma$-ray data may be achieved only for much closer FRBs to be detected in the future. Such significant constraints are illustrated in Figure 3, in which we assume a pseudo FRB with a redshift $z=0.01$.
One can see that the magnetar parameters are significantly constrained. This method can be readily applied to future FRBs that are localized to  much smaller distances than FRB 121102.

After the submission of this paper, we noticed that \cite{xi17} processed the LAT data of known FRBs and placed constraints on the possible prompt GRB-FRB associations. They did not consider the possibility of the birth of a magnetar before the FRBs, which is the subject of our paper.

\acknowledgments
BBZ acknowledges support from the Spanish Ministry Projects AYA 2012-39727-C03-01 and AYA2015-71718-R. The computation resources used in this work are owned by Scientist Support LLC. BZ acknowledges NASA NNX14AF85G and NNX15AK85G for support.

\begin{figure}

\begin{tabular}{c}
\includegraphics[keepaspectratio, clip, width=0.45\textwidth]{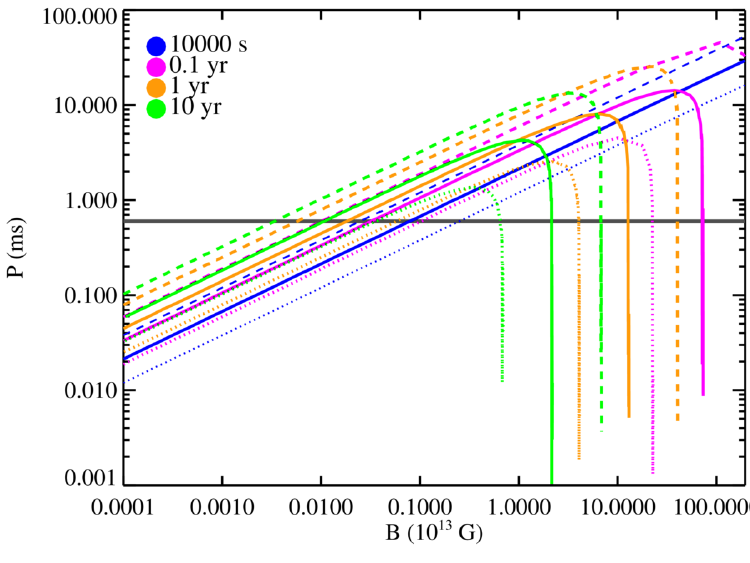} \\
\end{tabular}
\caption{Same as Figure 2 for a pseudo FRB at redshift $z=0.01$ with a same LAT flux limit as FRB 121102.}
\label{fig:fig3}
\end{figure}


\begin{thebibliography}{}
\expandafter\ifx\csname natexlab\endcsname\relax\def\natexlab#1{#1}\fi

\bibitem[{{Acero} {et~al.}(2015){Acero}, {Ackermann}, {Ajello}, {Albert},
  {Atwood}, {Axelsson}, {Baldini}, {Ballet}, {Barbiellini}, {Bastieri},
  {Belfiore}, {Bellazzini}, {Bissaldi}, {Blandford}, {Bloom}, {Bogart},
  {Bonino}, {Bottacini}, {Bregeon}, {Britto}, {Bruel}, {Buehler}, {Burnett},
  {Buson}, {Caliandro}, {Cameron}, {Caputo}, {Caragiulo}, {Caraveo},
  {Casandjian}, {Cavazzuti}, {Charles}, {Chaves}, {Chekhtman}, {Cheung},
  {Chiang}, {Chiaro}, {Ciprini}, {Claus}, {Cohen-Tanugi}, {Cominsky}, {Conrad},
  {Cutini}, {D'Ammando}, {de Angelis}, {DeKlotz}, {de Palma}, {Desiante},
  {Digel}, {Di Venere}, {Drell}, {Dubois}, {Dumora}, {Favuzzi}, {Fegan},
  {Ferrara}, {Finke}, {Franckowiak}, {Fukazawa}, {Funk}, {Fusco}, {Gargano},
  {Gasparrini}, {Giebels}, {Giglietto}, {Giommi}, {Giordano}, {Giroletti},
  {Glanzman}, {Godfrey}, {Grenier}, {Grondin}, {Grove}, {Guillemot}, {Guiriec},
  {Hadasch}, {Harding}, {Hays}, {Hewitt}, {Hill}, {Horan}, {Iafrate}, {Jogler},
  {J{\'o}hannesson}, {Johnson}, {Johnson}, {Johnson}, {Johnson}, {Kamae},
  {Kataoka}, {Katsuta}, {Kuss}, {La Mura}, {Landriu}, {Larsson}, {Latronico},
  {Lemoine-Goumard}, {Li}, {Li}, {Longo}, {Loparco}, {Lott}, {Lovellette},
  {Lubrano}, {Madejski}, {Massaro}, {Mayer}, {Mazziotta}, {McEnery},
  {Michelson}, {Mirabal}, {Mizuno}, {Moiseev}, {Mongelli}, {Monzani},
  {Morselli}, {Moskalenko}, {Murgia}, {Nuss}, {Ohno}, {Ohsugi}, {Omodei},
  {Orienti}, {Orlando}, {Ormes}, {Paneque}, {Panetta}, {Perkins},
  {Pesce-Rollins}, {Piron}, {Pivato}, {Porter}, {Racusin}, {Rando}, {Razzano},
  {Razzaque}, {Reimer}, {Reimer}, {Reposeur}, {Rochester}, {Romani},
  {Salvetti}, {S{\'a}nchez-Conde}, {Saz Parkinson}, {Schulz}, {Siskind},
  {Smith}, {Spada}, {Spandre}, {Spinelli}, {Stephens}, {Strong}, {Suson},
  {Takahashi}, {Takahashi}, {Tanaka}, {Thayer}, {Thayer}, {Thompson},
  {Tibaldo}, {Tibolla}, {Torres}, {Torresi}, {Tosti}, {Troja}, {Van Klaveren},
  {Vianello}, {Winer}, {Wood}, {Wood}, {Zimmer}, \& {Fermi-LAT
  Collaboration}}]{2015ApJS..218...23A}
{Acero}, F., {Ackermann}, M., {Ajello}, M., {et~al.} 2015, \apjs, 218, 23

\bibitem[Abdo et al.(2010)]{2010ApJ...725L..73A} Abdo, A.~A., Ackermann, M., Ajello, M., et al.\ 2010, \apjl, 725, L73 

\bibitem[Beloborodov(2013)]{beloborodov13} Beloborodov, A. M. 2013, \apj, 777, 114

\bibitem[{{Cao} {et~al.}(2017){Cao}, {Yu}, \& {Dai}}]{cao17}
{Cao}, X.-F., {Yu}, Y.-W., \& {Dai}, Z.-G. 2017, ArXiv e-prints,
  arXiv:1701.05482

\bibitem[{{Chatterjee} {et~al.}(2017){Chatterjee}, {Law}, {Wharton},
  {Burke-Spolaor}, {Hessels}, {Bower}, {Cordes}, {Tendulkar}, {Bassa},
  {Demorest}, {Butler}, {Seymour}, {Scholz}, {Abruzzo}, {Bogdanov}, {Kaspi},
  {Keimpema}, {Lazio}, {Marcote}, {McLaughlin}, {Paragi}, {Ransom}, {Rupen},
  {Spitler}, \& {van Langevelde}}]{chatterjee17}
{Chatterjee}, S., {Law}, C.~J., {Wharton}, R.~S., {et~al.} 2017, ArXiv
  e-prints, arXiv:1701.01098

\bibitem[Cheng \& Zhang(2001)]{2001ApJ...562..918C} Cheng, K.~S., \& Zhang, L.\ 2001, \apj, 562, 918 

\bibitem[Connor et al.(2016)]{connor16} Connor, L., Sievers, J., \& Pen, U.-L.\ 2016, \mnras, 458, L19 

\bibitem[Cordes et al.(2016)]{cordes16} Cordes, J.~M., \& Wasserman, I.\ 2016, \mnras, 457, 232





\bibitem[Dai et al.(2006)]{dai06} Dai, Z.~G., Wang, X.~Y., Wu, X.~F., \& Zhang, B.\ 2006, Science, 311, 1127 

\bibitem[Dai et al.(2016)]{2016ApJ...829...27D} Dai, Z.~G., Wang, J.~S., Wu, X.~F., \& Huang, Y.~F.\ 2016, \apj, 829, 27 




\bibitem[Dai et al.(2017)]{dai17} Dai, Z.~G., Wang, J.~S., \& Yu, Y.~W.\ 2017, \apjl, 838, L7 

\bibitem[Fan \& Xu(2006)]{fan06} Fan, Y.-Z., \& Xu, D.\ 2006, \mnras, 372, L19 


\bibitem[{{Feldman} \& {Cousins}(1998)}]{Feldman-Cousins}
{Feldman}, G.~J., \& {Cousins}, R.~D. 1998, \prd, 57, 3873

\bibitem[{{Kashiyama} \& {Murase}(2017)}]{kashiyama17}
{Kashiyama}, K., \& {Murase}, K. 2017, ArXiv e-prints, arXiv:1701.04815


\bibitem[Li et al.(2017)]{2017ApJ...835...30L} Li, J., Rea, N., Torres, D.~F., \& de O{\~n}a-Wilhelmi, E.\ 2017, \apj, 835, 30 


\bibitem[{{Lorimer} {et~al.}(2007){Lorimer}, {Bailes}, {McLaughlin},
  {Narkevic}, \& {Crawford}}]{lorimer07}
{Lorimer}, D.~R., {Bailes}, M., {McLaughlin}, M.~A., {Narkevic}, D.~J., \&
  {Crawford}, F. 2007, Science, 318, 777

\bibitem[{{Marcote} {et~al.}(2017){Marcote}, {Paragi}, {Hessels}, {Keimpema},
  {van Langevelde}, {Huang}, {Bassa}, {Bogdanov}, {Bower}, {Burke-Spolaor},
  {Butler}, {Campbell}, {Chatterjee}, {Cordes}, {Demorest}, {Garrett}, {Ghosh},
  {Kaspi}, {Law}, {Lazio}, {McLaughlin}, {Ransom}, {Salter}, {Scholz},
  {Seymour}, {Siemion}, {Spitler}, {Tendulkar}, \& {Wharton}}]{marcote17}
{Marcote}, B., {Paragi}, Z., {Hessels}, J.~W.~T., {et~al.} 2017, \apjl, 834, L8

\bibitem[Metzger et al.(2008)]{metzger08} Metzger, B.~D., Piro, A.~L., \& Quataert, E.\ 2008, \mnras, 390, 781 


\bibitem[{{Metzger} {et~al.}(2017){Metzger}, {Berger}, \&
  {Margalit}}]{metzger17}
{Metzger}, B.~D., {Berger}, E., \& {Margalit}, B. 2017, ArXiv e-prints,
  arXiv:1701.02370

\bibitem[{{Murase} {et~al.}(2016){Murase}, {Kashiyama}, \&
  {M{\'e}sz{\'a}ros}}]{murase16}
{Murase}, K., {Kashiyama}, K., \& {M{\'e}sz{\'a}ros}, P. 2016, \mnras, 461,
  1498

\bibitem[{{Petroff} {et~al.}(2015){Petroff}, {Bailes}, {Barr}, {Barsdell},
  {Bhat}, {Bian}, {Burke-Spolaor}, {Caleb}, {Champion}, {Chandra}, {Da Costa},
  {Delvaux}, {Flynn}, {Gehrels}, {Greiner}, {Jameson}, {Johnston}, {Kasliwal},
  {Keane}, {Keller}, {Kocz}, {Kramer}, {Leloudas}, {Malesani}, {Mulchaey},
  {Ng}, {Ofek}, {Perley}, {Possenti}, {Schmidt}, {Shen}, {Stappers},
  {Tisserand}, {van Straten}, \& {Wolf}}]{petroff15}
{Petroff}, E., {Bailes}, M., {Barr}, E.~D., {et~al.} 2015, \mnras, 447, 246

\bibitem[{{Roe} \& {Woodroofe}(1999)}]{Roe-Woodroofe}
{Roe}, B.~P., \& {Woodroofe}, M.~B. 1999, \prd, 60, 053009

\bibitem[Scholz et al.(2016)]{2016ApJ...833..177S} Scholz, P., Spitler, L.~G., Hessels, J.~W.~T., et al.\ 2016, \apj, 833, 177 

\bibitem[Shapiro \& Teukolsky(1983)]{shapiro83} Shapiro, S.~L., \& Teukolsky, S.~A., \ 1983, Black Holes, White Dwarfs,
and Neutron Stars: The Physics of Compact Objects (New York: Wiley)

\bibitem[Sun et al.(2017)]{sun17} Sun, H., Zhang, B., \& Gao, H.\ 2017, \apj, 835, 7 



\bibitem[{{Tendulkar} {et~al.}(2017){Tendulkar}, {Bassa}, {Cordes}, {Bower},
  {Law}, {Chatterjee}, {Adams}, {Bogdanov}, {Burke-Spolaor}, {Butler},
  {Demorest}, {Hessels}, {Kaspi}, {Lazio}, {Maddox}, {Marcote}, {McLaughlin},
  {Paragi}, {Ransom}, {Scholz}, {Seymour}, {Spitler}, {van Langevelde}, \&
  {Wharton}}]{tendulkar17}
{Tendulkar}, S.~P., {Bassa}, C.~G., {Cordes}, J.~M., {et~al.} 2017, \apjl, 834,
  L7

\bibitem[{{Thornton} {et~al.}(2013){Thornton}, {Stappers}, {Bailes},
  {Barsdell}, {Bates}, {Bhat}, {Burgay}, {Burke-Spolaor}, {Champion}, {Coster},
  {D'Amico}, {Jameson}, {Johnston}, {Keith}, {Kramer}, {Levin}, {Milia}, {Ng},
  {Possenti}, \& {van Straten}}]{thornton13}
{Thornton}, D., {Stappers}, B., {Bailes}, M., {et~al.} 2013, Science, 341, 53

\bibitem[Vink \& Kuiper(2006)]{2006MNRAS.370L..14V} Vink, J., \& Kuiper, L.\ 2006, \mnras, 370, L14
\bibitem[Xi et al.(2017)]{xi17} Xi, S.-Q., Tam, P.-H.~T., Peng, F.-K., \& Wang, X.-Y.\ 2017, arXiv:1705.03657 


\bibitem[{{Yang} {et~al.}(2016){Yang}, {Zhang}, \& {Dai}}]{yang16}
{Yang}, Y.-P., {Zhang}, B., \& {Dai}, Z.-G. 2016, \apjl, 819, L12

\bibitem[Zhang(2001)]{zhang01} Zhang, B.\ 2001, \apjl, 562, L59 

\bibitem[Zhang(2013)]{zhang13} Zhang, B.\ 2013, \apjl, 763, L22 


\bibitem[Zhang(2017)]{zhang17} Zhang, B.\ 2017, \apjl, 836, L32 

\bibitem[{{Zhang} \& {M{\'e}sz{\'a}ros}(2001)}]{zhangmeszaros01}
{Zhang}, B., \& {M{\'e}sz{\'a}ros}, P. 2001, \apjl, 552, L35

\bibitem[Zhang \& Cheng(2002)]{2002ApJ...579..716Z} Zhang, L., \& Cheng, K.~S.\ 2002, \apj, 579, 716 


\end{thebibliography}
\end{document}